# Are Condorcet and Minimax Voting Systems the Best?[1]

**Richard B. Darlington**
**Cornell University**

## Abstract

For decades, the minimax voting system was well known to experts on voting systems, but was not widely considered to be one of the best systems. But in recent years, two important experts, Nicolaus Tideman and Andrew Myers, have both recognized minimax as one of the best systems. I agree with that. This paper presents my own reasons for preferring minimax. The paper explicitly discusses about 20 systems. Computer simulations show minimax picking better winners than 11 other voting systems, beating all but one system by surprisingly wide margins.

I recommend a new version of minimax called minimax-TD. TD completely or partially eliminates four minimax anomalies, including the two worst ones. Three features in TD all lead it to pick better winners, on average, than other versions of minimax. Those features are: (a) Smith/minimax, (b) one particular definition of a candidate's "largest loss," and (c) a multi-stage tie-breaker.

Comments invited.
rbd1@cornell.edu

1. Many thanks to Nicolaus Tideman, Andrew Myers, Sharon Weinberg, Eduardo Marchena, my wife Betsy Darlington, and my daughter Lois Darlington, all of whom contributed many valuable suggestions.

# Table of Contents

## 1. Introduction and summary

Elections come in many varieties. A group may be electing a single chief executive, or all the members of a legislative body, or the members of a group which performs both those functions. Many private groups have separate votes for president, vice president, secretary, and treasurer. Professional groups often vote on new members, such as when a department of clinical psychology needs to hire a new specialist on schizophrenia, or an orchestra needs a new second clarinet. Fraternities, sororities, clubs, and residential communities may vote on all new group members.

Concerns about proportional representation arise especially when electing legislative bodies. And when electing single individuals, it sometimes seems clear that the opinions of some voters should count more than others. In our orchestra example, perhaps the opinions of the conductor and first clarinet should count more than others. Here we make no attempt to consider all these variations, and we focus on the problem of electing a chief executive who fairly represents all the voters. As in most elections, we assume there are several candidates.

This type of election has concerned dozens of writers over several centuries. More detailed but now-dated descriptions of the area appear in Tideman (2006), Saari (2001), and Felsenthal (2012). Recently, Tideman (2019) picked the Condorcet-Hare system as best, with minimax and Hare as close runners-up. The Condorcet Internet Voting Service (CIVS), at civs.cs.cornell.edu, has managed over 25,000 nongovernmental elections and polls since its creation in 2003 by computer scientist Andrew Myers. In 2017 Myers switched CIVS's default voting system from the Schulze system to minimax, citing an earlier version of this paper as the main reason for the switch, though he had never met me personally when he switched. Minimax was invented independently by Black (1958), Simpson (1969), and Kramer (1977). Black described the system (p. 175), but never advocated it. Minimax attracted little praise for decades, so these recent developments represent a new level of recognition for the system.

I advocate three modifications to what may be the most popular version of minimax:

•Section 4.1 tells why I recommend the Smith/minimax system, in which minimax is applied only within the Smith set of candidates. This is the smallest set in which every candidate in the set beats everyone outside the set, in majority-rules two-way races. As explained in Section 4.1, Smith/minimax is nearly as easy as classic minimax to implement, and nearly always picks the same winner, but completely eliminates the two most serious anomalies in classic minimax.

•In minimax we often compute the size of each candidate's largest loss (LL) in their two-way races against other candidates. Three ways of computing LL have been suggested. Section 7 explains why I prefer to define LL as a candidate's largest *margin* of defeat.

•Section 7 also describes a minimax tie-breaker which is computationally simple, breaks nearly all ties, and on average picks better winners than random tie-breakers.

I'm sorry my defense of minimax can't be briefer. There is no easy way to dismiss 20-odd systems all at once. This paper is designed to be useful to both beginners and experts in this area. The paper's sections are arranged in the order which I believe beginners will find most useful. But more knowledgeable readers can read Sections 3-8 in almost any order. Section 2 briefly describes 17 voting systems described by Felsenthal (2012), plus the Schulze system (https://arxiv.org/abs/1804.02973), the Condorcet-Hare system advocated by Tideman (2019), and the STAR system (starvoting.us). Section 3 summarizes the anomalies and weaknesses of many of these systems. Section 4 analyzes eight anomalies in minimax, arguing that there are good reasons to ignore all of them. Sections 3-4 show that even before we examine computer simulations, it is reasonable to conclude that minimax is the best of the systems we discuss.

Section 5 turns to computer simulations. It considers two different models of voter behavior, often called “valence” and “spatial.” In valence models, candidates differ on only a single dimension we might call “excellence.” In spatial models, political issues which divide voters are represented by spatial axes, each voter and each candidate appears as a dot in that space, and each voter is presumed to prefer the candidates closest to themselves. A large study of hundreds of real elections, by Tideman and Plassmann (2012), found that spatial models fitted that data far better than valence models did. Section 5 describes four spatial-model simulation studies, in which minimax outperformed all but one system by wide margins. Table 1 in Section 5.2.2 shows the substantial and consistent superiority of Condorcet-consistent systems (not just minimax) over other voting systems. Skeptical experts in voting theory are urged to read at least Sections 4 and 5.

Section 6 discusses minimax’s resistance to insincere strategic voting. Section 7 examines several variants and close relatives of minimax, and chooses as best the tie-breaker and the definition of LL described above. Section 8 describes in more detail some of the problems with other voting systems, including a good deal of original material. Section 8 was placed last mainly because it’s quite long and its main conclusions are mentioned in earlier sections. Section 8 does describe other minimax advantages involving simplicity, transparency, voter privacy, and resistance to various types of dishonest manipulation.

I do not claim to have established the superiority of minimax and other Condorcet-consistent systems for all time, beyond all possible future evidence and analysis. But I do claim that several very different lines of evidence and reasoning suggest this conclusion given our current knowledge.

## 2. The variety of voting systems

This section says a little about 17 voting systems analyzed by Felsenthal (2012), plus three newer systems. Nearly everyone has used the very simple **plurality** (vote for one) system. Another system with very simple ballots is **approval voting**, in which each voter approves as many candidates as they wish, and the candidate with the most approvals wins. Also simple and moderately familiar is **plurality runoff**, which uses plurality voting followed by a runoff election between the top two candidates.

Several other systems use **ranked-choice ballots**, in which voters rank the candidates. Best-known of these is the **Hare** system, which is also called instant runoff voting (**IRV**) or single transferable vote (**STV**). In that system the candidate with the fewest first-place ranks is removed, and the ranks of all other candidates are recomputed as if that one candidate had never run. That process is repeated until only one candidate is left. The Hare system is often called “ranked-choice voting” (RCV), but it is actually just one of several systems using ranked-choice ballots. The **Coombs** system is like Hare, except the candidate removed in each round is the one with the most last-place ranks rather than the one with the fewest first-place ranks.

Some ranked-choice systems are **positional** systems, meaning that the winner is determined by the ranks received by each candidate. Two positional systems are especially well known. In the **Borda** system, we find for each candidate X the number of candidates ranked below X by each voter. Summing these values across voters gives each candidate's Borda count. The one with the highest Borda count is the winner. Thus, if there are no tied or missing ranks, the Borda winner is the candidate with the highest mean rank. In the **Bucklin** system a candidate wins if they receive a majority of the first-place ranks. If there is no winner by that rule, each candidate's number of "high ranks" is expanded to include second-place ranks, and any candidate who receives "high ranks" from over half the voters is the winner. If there is still no winner, "high ranks" are expanded to include third-place ranks. And so on. If two or more candidates all receive "high ranks" from over half the voters, the one with the most "high ranks" is the winner.

Other systems use rating scales. In **range voting**, each voter rates each candidate on a multi-point scale, often ranging from 0 to 100, and the candidate with the highest mean rating wins. The **majority judgment** system also uses a rating scale, though the points on the scale have non-numeric verbal labels like "excellent" and "poor," and the number of points on the scale is much smaller – usually 5 to 8. The winner is the candidate with the highest median rating. A tie-breaker described in Section 8.3.1 breaks nearly all ties. These two systems are the best-known rating-scale systems.

If ballots use ranks or rating scales, election managers can use the ballots to run a two-way race between each pair of candidates. If a candidate wins all their two-way races by majority rule, they are called a **Condorcet winner**. A voting system has **Condorcet consistency** (CC) if any Condorcet winner is named the winner. Surprisingly, none of the voting systems just described has CC, although all of them seem reasonable at first glance.

A **Condorcet paradox** occurs if there is no Condorcet winner. There is then a **Condorcet cycle** of three or more candidates. In a typical cycle, A beats B by majority rule, and B beats C, but C beats A. In the simplest possible example, one voter ranks three candidates as ABC, one as BCA, and one as CAB. Then A beats B 2:1, B beats C, and C beats A. Large studies by Gehrlein (2006) and Tideman and Plassmann (2012) both found that Condorcet paradoxes occur only occasionally in real data, but they are common enough that it's important to handle them as well as possible.

CC systems typically allow tied ranks. If a voter fails to rank a candidate, they are typically presumed to rank them below anyone whom they did rank explicitly. I agree with those rules. All these systems name any Condorcet winner as winner, so we describe only what they do under a Condorcet paradox. In that event the **Black** system names the Borda winner as winner. The **Copeland** winner is the candidate who wins the largest number of their two-way races. It was recently discovered that this system had been proposed by 1283, for electing church officials, by Cardinal Ramon Llull; see Colomer (2013). In the **Young** system, we find for each candidate X the smallest number of voters who would have to be removed to turn X into a Condorcet winner, and the winner is the one for whom that number is smallest. In the **Dodgson** system, an "interchange" occurs if a voter switches two candidates whom they had ranked adjacently, as when a voter changes the ABCD ranking to ACBD or ABDC. The Dodgson winner is the candidate who would need the fewest interchanges, summed across voters, to become a Condorcet winner. In the **Nanson** system, all candidates with Borda counts at or below the mean are removed and Borda counts are then recomputed. That process is repeated until one candidate has a majority of first-place ranks. Nanson is the only system mentioned here which has CC even though it never explicitly examines the results of all possible two-way races. That's because a Condorcet winner doesn't always have the highest of all Borda counts, but it never has a Borda count at or below the mean of those counts, and thus is never eliminated by Nanson.

**Minimax** has CC. If there is no Condorcet winner, we find each candidate's largest loss (LL) in their two-way races. The candidate with the smallest LL is the winner. I recommend a particular version of minimax I call minimax-TD for minimax-Tideman-Darlington. TD differs from other versions of minimax in three main ways. First, it uses Smith/minimax, as recommended to me by Nicolaus Tideman. The Smith set is the smallest set of candidates in which every candidate in the set beats every candidate outside the set in a majority-rules two-way race. TD picks its winner from within the Smith set. That set is easy to identify, because every member of the Smith set wins more two-way races than anyone outside the set. Thus, you can start with an empty set, rank the candidates by the number of races they won, and add candidates to the set in that order until everyone in the set beats everyone outside.

TD's second difference from other versions of minimax is that it defines candidate X's value of LL as X's largest *margin* of loss. Other versions of minimax define LL as the largest number of voters who voted against X in any race which X lost, or as the largest number who voted against X in any race at all. Section 7 explains why I prefer the definition using margins. TD's third difference is a new tie-breaking system for minimax which is simple and intuitively reasonable, breaks nearly all ties, and is shown by simulations to pick better winners than random tie-breakers. It too is explained in Section 7.

All the CC systems just mentioned (including minimax but not TD) are described by Felsenthal (2012). So far, we have named 9 voting systems without CC, and 6 with CC. The **Kemeny** system is a CC system described by Felsenthal, but little used because its computations are extremely complex with larger numbers of candidates. The **Schulze** system is used fairly widely by private organizations, and is studied in Section 5.2, though its procedures are more complex than I wish to explain here. The **Condorcet-Hare** system was not mentioned by Felsenthal, but was recommended by Tideman (2019) and is discussed in Sections 3, 5, and 6. The simplest version of this system picks the Condorcet winner if there is one, and picks the Hare winner otherwise.

The computer simulations in Section 5.2 also include the STAR voting system, found on the internet at starvoting.us. In this system, each voter rates all candidates on a scale, using the integers from 0 to 5 with 5 high. To lessen the effects of insincere strategic voting, each voter must use the entire scale, rating at least one candidate 0 and at least one 5. The system then takes the two candidates with the highest mean ratings, and automatically performs a runoff election between those two, counting for each one the number of voters who rated them above their opponent.

## 3. Some electoral criteria violated by minimax's competitors

A voting system is said to suffer an *anomaly* or *paradox* if an artificial example can be created which demonstrates some clearly undesirable property of the system, such as if giving candidate X a higher rank can sometimes make X *lose*. An *electoral criterion* is a rule requiring the absence of a particular anomaly. Electoral theorists have widely agreed for decades that every known voting system, and indeed every possible system, suffers from at least one anomaly, and often from several. Felsenthal (2012) summarized this literature, describing 16 electoral criteria and showing that every one of his 18 systems violates at least six of those criteria. For those interested, three other books by Felsenthal and Hannu Nurmi, dated 2017, 2018, and 2019, probe electoral anomalies even more deeply. The inevitable conclusion from all this is that some anomalies must be ignored. Section 4 explains why we can dismiss all eight of the anomalies violated by minimax. This section describes some other criteria which I favor keeping.

It is widely agreed that most anomalies occur only rarely, if ever, in real data. I'll distinguish anomalies from *weaknesses* or *problems*, which may occur more frequently. For instance, vote-splitting is a ubiquitous problem in plurality elections with more than two candidates; candidates A and B might both be able to beat C in two-way races, but they both lose to C in a plurality election because they split their vote. Of the items discussed in the rest of this section, I would classify most as "weaknesses" because there is little serious doubt about their potential frequency. But the first item (nonmonotonicity) is more of an anomaly.

**Monotonicity.** Felsenthal and Nurmi (2017) devote an entire book to monotonicity. Section 1.2 of that book makes clear that one cannot even give an exact definition of monotonicity without using a lot of highly technical language. There are competing conceptions of monotonicity, and even Felsenthal and Nurmi distinguish between two different types of monotonicity. Since this paper is designed to be read by nonspecialists, I will give only a simple approximate definition of the concept. Define a simple vote change as a change by one or more voters who all vote identically and then all change their votes identically. A voting system is nonmonotonic if a simple vote change ever changes the election's outcome in the opposite direction from the obvious intention of those voters. Thus, a system is nonmonotonic if any of these outcomes could ever occur:

1. Candidate X changes from a winner to a loser when a simple vote change raises X's ranking without changing the ranks of other candidates relative to each other.
2. X changes from a loser to a winner when the change lowers X's ranking without changing the ranks of other candidates relative to each other.
3. X changes from a winner to a loser when new identical voters appear, all ranking X first.
4. X changes from a loser to a winner when new identical voters appear, all ranking X last.

Felsenthal and Nurmi (2017) do give a simple and clear statement on page 11: of the 18 voting systems they are considering – the same systems discussed by Felsenthal (2012) – the only ones which satisfy both of their conceptions of monotonicity are plurality, approval, Borda, range voting, and minimax. Monotonicity seems like a particularly important condition, since the charge that a voting system is nonmonotonic could be used by losers who want to challenge election results and perhaps undermine democracy by challenging elections in general.

**Strategic voting.** The most common form of strategic voting is to give an insincerely low rating or ranking to a candidate X whom you consider moderately attractive, to prevent X from beating other candidates you favor even more. Or a voter might rate or rank some other candidate Y insincerely high, because they believe Y has the best chance of beating other candidates whom the voter dislikes more than Y. Section 8 argues that the danger of strategic voting cannot be ignored, and that voting systems do differ substantially in resistance to that danger. The range voting, majority judgment, Borda, Coombs, and Dodgson systems appear to be particularly susceptible. Green-Armytage, Tideman, and Cosman (2016) studied the problem at length, and concluded that the Condorcet-Hare system has the best combination of efficiency (ability to pick the best winners) and resistance to strategic voting, with minimax coming in second. My simulations in Section 6 agree with Green-Armytage, Tideman, and Cosman that Condorcet-Hare is theoretically far superior to minimax on strategy resistance, but those in Section 5.2 conclude that the reverse is true on efficiency. Section 6 also argues that, when one considers all the real-life complications and dangers faced by strategic voters under either of these two systems, it seems unlikely that strategic voting will be a major problem under either of these systems.

Therefore, once we have narrowed the choice to these two systems, we can ignore the problem and use the most efficient system, which is minimax.

**Completeness.** I'll call a ballot "incomplete" if it doesn't allow a voter to express a preference between every pair of candidates. Among the systems considered here, the only ones suffering this problem are plurality, approval, and plurality with runoff. For instance, suppose in an approval election a voter would rank the candidates ABCD, and that voter chooses to approve A and B. But if it later turns out that the top two candidates are A and B, or are C and D, the voter has in effect been prevented from voting in the two-way race which determines the final winner. Incomplete ballots are often assumed to make voting easier, but Section 8.1 explains why that's not so. Thus, those ballots are unacceptable. Incomplete systems were consistently among the worst performers in the simulation studies in Section 5.2.

Of the voting systems discussed by Felsenthal (2012) and described in Section 2, minimax is the only one passing all of the three criteria discussed so far. But many systems also fail other criteria named below.

**Simplicity.** A system fails this criterion if the average citizen would find it difficult even to understand the system or explain it to others. In my opinion the well-known Kemeny and Schulze systems clearly fail this criterion. As mentioned in Section 2, the *goals* of the Dodgson and Young systems are easily described. But papers cited in Section 8.4 show their computations can be a challenge even for modern computers. Thus, these two systems also fail our simplicity criterion.

**Ease of voting.** Plurality and approval ballots are often assumed to make voting easier than ranked-choice or rating-scale ballots. But plurality ballots certainly aren't easy for a voter who prefers some candidates over others but likes two or more top candidates equally well. And Section 8.1 explains why approval ballots can actually make voting very difficult for the voter who wants to maximize their influence, as most voters do. But ranked-choice ballots can make voting surprisingly easy if tied and missing ranks are allowed. Then voting will be even easier than plurality voting for anyone who likes two or more top candidates equally well, and also easier than approval voting (see Section 8.1). If there are many candidates, voting may be quite difficult with rating scales or with ranked-choice ballots if tied and missing ranks are not allowed. That's often the case with the Hare and Coombs systems, since some versions of those systems must know at each stage of the elimination process which one candidate is each voter's top or bottom choice among those remaining. CC systems allow tied and missing ranks, so voting may be easier with them than with many other systems.

**Resistance to vote-splitting and spoiling.** These problems arise primarily with plurality voting and with Hare. In plurality voting, the vote-splitting problem arises if two or more very popular candidates are so similar to each other that they all lose because they split the votes of the voters attracted to those candidates. The similar "spoiler" problem arises if there is one candidate X who could beat anyone else in a two-way race, but one or more less-popular candidates take some of X's votes so that X loses to some other candidate. The next paragraph shows how Hare shares these same problems.

**Straddling.** I made up this term to describe a weakness I believe exists in the Hare and Condorcet-Hare systems. Imagine a candidate X who could beat any other candidate in a two-way race, but who is surrounded by other fairly similar candidates, some more liberal than X and some more conservative. X may then get very few top ranks in ranking systems, and thus be removed early by the Hare system. The Condorcet-Hare system was invented to avoid this precise problem, but that system still has an unfortunate tendency to remove the "best" candidates as defined in Section 5.1.

**Condorcet consistency (CC).** Several different lines of reasoning can lead to the conclusion that CC is essential. Many people simply consider the point to be self-evident. Others may be persuaded by the fact that all non-CC systems are eliminated by the needs for monotonicity, completeness, and reasonable resistance to burying. Still others are persuaded by the argument that CC is politically necessary just because so many ordinary citizens consider it important. This was illustrated by the 2009 mayoral race in Burlington, VT, a liberal university town which had recently adopted the Hare system. In that race, Hare eliminated the Democratic candidate before either the Republican or Progressive candidate (who won), even though the Democratic candidate was the Condorcet winner, beating each of five other candidates in two-way races. Voters found that bizarre and promptly voted to abandon the Hare system. Section 8.5 explains how this can happen, and why we might expect to see it fairly often.

The Condorcet Internet Voting Service (CIVS), at civs.cs.cornell.edu, is the only free, highly secure, open-source voting service managed by a prominent researcher – Andrew Myers, who is also editor-in-chief of TOPLAS, the premier journal on computing languages. CIVS offers users the choice of several CC voting systems, some of which are newer and lesser known than the systems listed in Felsenthal (2012).

## 4. Dismissing eight criteria violated by minimax

This section describes the eight electoral criteria which Felsenthal (2012) identifies as being violated by minimax, and explains why all those violations can be ignored. Much of the material in previous sections will be familiar to experts in voting theory, but most of the material in Section 4 is original.

### 4.1 The absolute loser, Condorcet loser, and preference inversion criteria

Of the 16 electoral criteria described by Felsenthal (2012), five were identified in that chapter as criteria whose violation is widely regarded as "especially intolerable." One of these is the **absolute loser criterion**, which states that no candidate should ever win if they are ranked last by over half the voters. Another is the **Condorcet loser criterion**, which says that no candidate should ever win if they lose all their two-way majority-rule races against other candidates. Any absolute loser is a Condorcet loser, since anyone ranked last by most voters will lose all their two-way races.

Artificial Example 3.5.11.1, on pages 61-62 of Felsenthal (2012), shows that minimax can violate both of these criteria. In that example, three candidates A, B, C are locked in a Condorcet cycle, with large margins of defeat in all three of the elections between any two of them. D is ranked first by just under half the voters, and last by just over half. Thus, D loses to all other candidates by very small margins, and is thus the minimax winner despite being the Condorcet and absolute loser. These two violations led Felsenthal (2012) to dismiss minimax. But no absolute loser or Condorcet loser is ever in the Smith set, and therefore never wins in Smith/minimax. So, Smith/minimax satisfies these criteria.

Two computer simulations confirm both the high similarity between Smith/minimax and classic minimax, and the superiority of Smith/minimax when they do differ. The simulations used the general approach described in Section 5. Data sets generated in this way are much more likely to produce anomalies like this when there are only few voters per trial, so I used very small numbers to find more of these anomalies. With 4 candidates and only 5 voters per trial, it took just over a million trials to find 1000 trials in which the two systems picked different winners. Smith/minimax picked the better winner in 849 of those trials. With 4 candidates and 25 voters per trial, it took almost 46 million trials to find 1000 trials with different winners, and Smith/minimax picked the better winner in 752 of those trials. I assume that with more voters per trial, the two versions would become even more similar, both in the rarity of differences between them, and in the quality of the winners picked.

A voting system violates the **preference inversion criterion** if it would ever pick the same winner if every voter's ranks were reversed. If all ranks were reversed in the Felsenthal example cited above, D would be an absolute winner (ranked first by over half the voters) and would win under almost all voting systems including minimax. Thus, classic minimax also violates the preference reversal criterion. Smith/minimax eliminates any violation of this criterion which occurred because a Condorcet loser or absolute loser had been picked as winner, but Markus Schulze pointed out to me that minimax can violate the preference inversion criterion even without those other violations. But this appears only very rarely in computer simulations with thousands or millions of trials. In these cases, one candidate won or lost all their races by small margins, while all other candidates lost at least one race by a larger margin, both before and after inversion. Sometimes the minimax winner in these cases was actually the best candidate in having the smallest mean distance from the voters. Thus, it's not completely clear that this should even be called an "anomaly." After all, anomalies are supposed to illustrate a system's faults, and a system is not at fault if it picks the best winner.

### 4.2 Three anti-manipulation criteria

This section considers three other electoral criteria from Felsenthal's list of eight criteria violated by minimax. Felsenthal calls them the no-show, twin, and truncation criteria. I'll call them "anti-manipulation" criteria, since they are all intended to prevent various kinds of dishonest electoral manipulation. A voting system violates the no-show criterion if a voter can sometimes benefit by abstaining rather than voting sincerely. The twin criterion is very similar: it prohibits an anomaly in which two people who rank candidates the same find it beneficial for one of them to vote and for the other to abstain. The truncation criterion is also somewhat similar: it is violated if a voter can benefit by ranking only their first few choices rather than all of them.

Felsenthal (2012, p. 63) presents a 19-voter artificial example in which 5 voters rank four candidates in the order DBCA, 4 others rank them BCAD, 3 others ADCB, 3 others ADBC, and 4 others CABD. He credits the example to Hannu Nurmi. The minimax winner here is B. In the no-show and twin anomalies, three voters from the last pattern choose to abstain. This changes the winner to A, whom those voters prefer to B, thus violating those criteria. In the truncation anomaly, all four voters in the last pattern truncate their ballots, giving only their first two choices. This changes the minimax winner to C, who is their first choice. So, minimax violates the truncation criterion.

These are interesting as mathematical curiosities, but there are seven reasons I find them unconvincing as examples of possible real-world manipulations. *First*, the data contain two Condorcet cycles: one with candidates A, B, and C, and one with A, C, and D. But even one cycle is quite rare in real data. *Second*, all the manipulations took 3 or 4 of the 19 voters, yet the key margins of victory (both before and after the manipulations) were all by one vote each. Thus, there is nothing like our examples of burying or max-and-min in Sections 8.2.1 and 8.3.2, in which a few strategic voters reverse very large margins of victory. *Third*, to know their maneuvers would benefit them, the strategic voters would have to know almost exactly how everyone else would vote, including the two Condorcet cycles. But such certainty is rare; recall that on US Election Day in November 2016, some respected analysts were saying with 99% confidence that Hillary Clinton would win the presidency. As we'll see in Section 8, burying and max-and-min may require no such knowledge. *Fourth*, if the strategic voters guess wrong, they would very likely harm themselves, since they are withholding all or part of their votes. *Fifth*, the voters must be persuaded to actually execute this complex and counter-intuitive plan. It makes me imagine a college math class with a full blackboard. *Sixth*, in most real elections there would have to be hundreds or thousands of strategic voters to whom this complex plan must be explained. The plan would surely end up in local newspapers, eliciting a mixture of outrage, amazement, and ridicule. This would alienate some voters who had previously planned to vote for the candidate benefited by the scheme, and it would stimulate opposing voters to actually vote. *Seventh*, after all that effort and risk, in two of the three maneuvers the participating voters don't even get their favorite candidate, just their second-favorite one. For all these reasons, this example belongs in *Alice in Wonderland*; it does not seem like a maneuver anyone would actually try to execute in real life.

The electoral criteria discussed in Sections 4.1, 4.3, and 4.4 are intended to ensure selection of the best candidates, but are not designed to thwart dishonest manipulations. So, I'll call them *optimizing criteria* in contrast to the anti-manipulation criteria of the current section. The anti-manipulation criteria are also intended to ensure selection of the best candidates even in the absence of attempts at dishonest manipulation, so they too are optimizing criteria in a sense, and we should consider that possibility. But the whole purpose of optimizing criteria is to promote selection of the best candidates. Thus, when one optimizing criterion conflicts with another (and these all conflict with CC), it seems reasonable to use computer simulations to see which criteria lead to selection of the best candidates. Section 5 describes four simulation studies which found that the best candidates were selected by minimax, which violates all of the criteria in Section 4. And we see in the next two subsections that there are also other reasons for dismissing the criteria discussed there.

## 4.3 SCC/IIA

Minimax violates what Arrow (1963) called independence of irrelevant alternatives (IIA) and Felsenthal (2012) called the subset choice criterion (SCC). A voting system violates IIA if removing a loser (the "irrelevant alternative") from the contest can change the winner. Arrow considered the need for IIA essentially self-evident (p. 26), though Tideman (2006, p. 132) disagreed. In their later book introducing and defending the majority judgment (MJ) voting system, Balinski and Laraki (2010, p. 52) vigorously defended the need for IIA, and regarded MJ's compliance with IIA as a major argument for that system. That book has been praised by several important figures, so we need to examine IIA more closely. There are several arguments against it.

First, suppose there is a Condorcet cycle in a three-way race, so each candidate wins one two-way race and loses one. A voting system picks a winner, but then the candidate beaten by that winner drops out, leaving only that winner and the candidate who beat them by majority rule (MR). If the system satisfies IIA by keeping the same winner, then it violates MR in a two-way race. Only two major voting systems (MJ and range voting) don't reduce to MR in two-way races. Those are also the only two major systems which comply with IIA. Both those systems use rating scales. In range voting the winner is the candidate with the highest mean rating; in MJ it's the candidate with the highest median. All the other major anomalies in voting systems arise only with three or more candidates, and nearly all experts agree that with only two candidates, MR is obviously the best system. Also, neither MJ nor range voting has Condorcet consistency, so we can regard IIA as conflicting with that criterion.

Second, MJ performs very poorly in the computer simulations in Section 5.2.2.

Third, it's not clear that violation of IIA is actually a flaw. Suppose we were planning a plurality election, and we feared that some candidates might drop out after votes are cast. To deal with that, we might allow each voter to rank their top several choices, so vote-counters would know each voter's top choice even after some candidates drop out. That clearly assesses voter desires better than the obvious alternative, which would be to ignore the votes of those who had voted for dropouts. But that latter system satisfies IIA, whereas our improvement makes the system violate IIA. For instance, suppose A would win with no dropouts, but B drops out and most of B's votes go to C, who then beats A. So, the superior system violates IIA and the inferior system satisfies it.

Fourth, if IIA is self-evident, it should also apply to choosing champions in sporting contests. But imagine a sports league with three teams in which each pair of teams plays 9 games. Team A has won all 9 of its games against B, and B has won all its games against C, but C has beaten A 5 games to 4. Thus, A has won 13 games, B has won 9, and C has won 5. If we named A the league champion, B would be a loser. But if B were suspended for hazing and became ineligible, and we therefore ignored the results of B's games, we would be forced to choose C, who had beaten A 5 games to 4. Thus, the initial choice of A violates IIA. But clearly team B's ineligibility doesn't mean that the results of its games are irrelevant to the choice between A and C. We thus see that even after a team becomes ineligible to win, the *data* generated by the team's candidacy may be useful in choosing a winner from the other teams. The problem with IIA is even worse, because even if B isn't suspended or even accused of wrongdoing, we know that B *could* be suspended, so that choosing A would violate IIA.

This sports example is not directly analogous to elections, so Darlington (2017) ran a series of computer simulations of elections, to see whether the data generated by losers would be useful in choosing among the remaining candidates. He used spatial models like those described in Section 5.1. In these models, the computer "knows" the "true excellence" of each candidate, and can thus tell which of several voting systems picks the best candidates on average. Because of the close link Balinski and Laraki (2010, p. 52) made between IIA and MJ, Darlington identified the MJ winner in each trial, removed that candidate and found the MJ winner from among the remaining candidates. MJ clearly considers the first of those two candidates to be the best choice as overall winner. Advocates of IIA have already concluded that the best system using "relevant" data is MJ, so we should see whether any system using "irrelevant" data can improve on MJ. Darlington (2017) devised two different voting rules using "irrelevant" data. To make any rejection of IIA as difficult and convincing as possible, he created rules which use *only* data which IIA calls irrelevant. He compared each of these two rules to MJ under 12 different conditions, making 24 comparisons between MJ and systems violating IIA. Space prevents a full description here of the two voting rules and 12 conditions. Each of these 24 comparisons was made in one million separate trials. Then, in the trials in which the two voting systems picked different winners, the computer printed the percentage of those trials in which the rule using only "irrelevant" data had picked a better winner than MJ had picked. These 24 values are on lines 4-5 of Table 1 in Darlington (2017). All but three of those 24 values were over 50; their mean was 58.25, their median was 59.5, and their highest value was 71. Thus, we see that voting systems using *only* "irrelevant" data can choose between two candidates far more accurately than MJ itself can. It's hard to imagine a more convincing repudiation of IIA.

### 4.4 Multiple districts

The multiple-districts criterion is violated if a candidate could win in each of two or more districts but lose if all the districts are merged into one. I know of two examples showing that classic minimax fails that criterion. One is Example 3.5.11.4 in Felsenthal (2012, p. 64), the other is in Darlington (2016, Table 3, p. 13). In both of those examples, minimax picks the absolute loser as winner. Since minimax-TD never does that, I know of no examples in which minimax-TD fails the multiple-districts criterion.

In summary, Section 4 includes arguments for dismissing all eight of the electoral criteria listed by Felsenthal (2012) as being violated by minimax.

## 5. Simulation studies on voting systems

### 5.1. Why our computer simulations use spatial models of voter behavior

This paragraph summarizes Section 5.1 so readers can decide whether they can skip to Section 5.2. Voters may respond to differences among candidates on policy dimensions like liberalism-conservatism, or on general-excellence traits like energy, honesty, intelligence, and experience, or on both excellence and policy dimensions. Pure spatial models (explained shortly) recognize differences on policy but not on excellence. But differences among candidates in excellence tend to suppress Condorcet paradoxes while policy differences on multiple issues tend to produce them. Previous sections presented several reasons for dismissing non-CC voting systems: flaws involving burying, incompleteness, vote-splitting and spoiling, and the rejection of non-CC systems by both the general public and many electoral theorists. The obvious major problem faced by CC systems is the Condorcet paradox. Thus we want to understand how Condorcet paradoxes arise, and also need to study how various voting systems behave under Condorcet paradoxes. Using a voting model which suppresses Condorcet paradoxes would interfere with both of those goals. Therefore our simulations should use spatial models.

We now explain these points in more detail. Some models assume that voters all have the same goals, but they disagree on which candidates would pursue those goals most effectively, so the election's primary purpose is to maximize attainment of the agreed-on goals. This case might arise when the members of a nonprofit organization are voting to choose the organization's president, and they all agree on the organization's goals. The directors of a profit-making firm may also all agree on the firm's goals. In such models, the only relevant differences among candidates are on a trait we might call "excellence" or "general attractiveness," and voters disagree with each other only because of random differences in their perceptions of each candidate's true score on that trait.

Other models assume that voters have conflicting goals, so the election's primary purpose is to compromise among those competing goals. This case would presumably arise more often in public elections, where voters may differ on the desired amounts of military spending, business regulation, social-welfare spending, tax policies, regulations on drugs and sexual behavior, and other issues. This case is usually studied with spatial models. In these models we treat each area of disagreement (tax policy, military spending, etc.) as a policy dimension. Each dimension is represented as an axis in space, and we represent each voter and each candidate as a dot in that space, according to their positions on those issues. Each voter is presumed to rank the candidates by their distance from themselves, with the closest candidate ranked highest. In the simplest possible spatial model, voters and candidates are assumed to differ on just one policy dimension, frequently labeled liberal-conservative or left-right. That model should not be confused with the excellence model of the previous paragraph, which has no policy dimensions at all.

If we assume that voters are mutually independent and that we can measure distortions from sampling error, it's quite easy to tell whether voters disagree on policy dimensions. Suppose first that we have at least four candidates. If voters all have the same goals, a voter's choice between two candidates A and B should not correlate significantly with the choice between two other candidates C and D, because all differences among voters are presumed to be produced by random errors of judgment, which are presumed to be mutually independent. But such correlations could easily occur under a spatial model, even a model with just one policy dimension, as when A and C are both more liberal than B and D.

A different test must be used if there are only three candidates. Under an excellence model, differences among voters concerning the excellence of one candidate B will tend to produce a positive correlation between the preference for B over another candidate A and the preference for B over a third candidate C; but such correlations should never be more negative than the laws of chance would allow. But under a spatial model, suppose those candidates fall in the order A B C from political left to political right. Then those on the left will prefer A to B and those on the right will prefer C to B, but few if any will prefer both A and C to B, so the aforementioned correlation will be negative. If any three candidates all differ on a left-right dimension, as in this example, there will always be one candidate between the other two, so negative correlations like this should appear. This is shown by an artificial-data example I ran with 1000 voters, 50 candidates, and two policy dimensions, with both voters and candidates drawn randomly from a bivariate normal distribution. In samples of 1000, all correlations below -0.1 are significant beyond the 0.001 level one-tailed. With 50 candidates the number of ABC correlations of this sort is 50·49·48/2 or 58,802. In my sample, 4075 of those 58,802 correlations were below -0.1, and the most negative correlation was -0.754.

Tideman and Plassmann (2012) studied real-data elections with several candidates. From each election they formed all possible three-candidate sets, and tested whether those three candidates seemed to differ on policy dimensions. After studying hundreds of these sets, they concluded that the evidence was overwhelming that policy dimensions were important in the elections they studied. Many voting systems, including the famous Borda and Kemeny systems, were specifically developed for excellence models. That suggests these systems will perform poorly in computer simulations using spatial models – a suspicion confirmed by a whole series of computer simulations by Darlington (2016).

It seems plausible *a priori* that the candidates in a set might differ from each other on both excellence and policy dimensions, and the results of any of the aforementioned analyses would not eliminate that possibility. But intuition suggests that the larger the differences are among candidates on general excellence, the less likely a Condorcet paradox is to appear. My own unpublished computer simulations confirm this conclusion. Thus differences in excellence don't produce Condorcet paradoxes, but rather prevent them. Therefore they don't help us understand how the paradoxes arise, and don't help us see which voting systems behave best under those paradoxes. Thus our computer simulations should use pure spatial models.

Before asking *when* a spatial model could produce a Condorcet paradox, we'll see how it could *ever* do so. For a simple example, suppose there are two policy dimensions X and Y. Suppose three candidates A, B, C are respectively at (1,4), (5,5), and (6,1) on X and Y, and three voters R, S, T are at (2,2), (3,6), and (7,3) respectively. Then the voter-candidate Euclidian distances in the XY space are

| | R | S | T |
|---|---|---|---|
| A | 2.24 | 2.83 | 6.08 |
| B | 4.24 | 2.24 | 2.83 |
| C | 4.12 | 5.83 | 2.24 |

Thus voter R will rank the candidates ACB, S will rank them BAC, and T will rank them CBA, producing a Condorcet paradox. Many similar paradoxes could be created, with either few or many voters.

## 5.2 Four computer simulations

### 5.2.1 Features and purposes of these studies

The simulation studies in this section contain five features designed to increase their sensitivity to differences among voting systems. First, each study included 10 artificial candidates – more than in many studies. This increased the number of times any two systems would pick different winners, thus increasing the sensitivity to differences among systems. Second, one system was counted as "beating" another in a trial if it merely picked a "better" candidate (defined later) in that trial. That produced much larger numerical differences among systems than if we merely counted differences in picking the "best" candidate. Third, when comparing two systems, the trials the better system won were expressed as a percentage of the trials in which those two systems had picked different winners. If two systems rarely picked different winners, but one of them picked the better winner in a large majority of those rare trials, I believe most people would consider that an important difference, especially if the winning system is the simpler of the two systems, which minimax often is. Fourth, trials with and without Condorcet winners were analyzed separately, since differences among systems might appear only or mainly in one of those cases. Fifth, three of the four studies included 100,000 trials – more than are often used. These design features may explain why I found larger differences among systems than have been found in many other studies.

In each trial, both candidates and voters were drawn randomly and independently from a standardized bivariate normal distribution with mutually independent variables. Each voter's rating of each candidate was defined as 10 minus the Euclidean distance between the two. Each candidate's "true excellence" was defined as that candidate's mean of those ratings. These ratings were also used to determine each voter's voting pattern, which then determined each voting system's winner.

Ties between candidates are a problem. Many important elections have many thousands or even millions of voters. But simulation studies need to run many trials, and it's not practical to have so many voters in each of so many trials. But if we use fewer voters in each trial, ties may arise frequently, and some voting systems handle ties better than other systems. We don't want to bias our results against those systems, since ties are so rare in larger elections. To handle that problem, these studies excluded from the entire analysis all trials in which any of the studied systems yielded a tie. I used a version of minimax, described in Section 7, which breaks nearly all ties, so this exclusion of trials with ties eliminated a possible bias favoring minimax.

In computer simulations, it may be that some voting systems lose more accuracy in small samples than other systems do. I chose to study that by sometimes using just 100 voters per trial and other times using 1000. If one system's advantage over others faded with larger samples, this approach would help reveal that.

### 5.2.2 Further description of the studies

The version of minimax used in these studies did not include the restriction to Smith-set candidates described in Section 4.1. The computer simulations in that section suggest that this fact biased the results of these studies *against* minimax, though probably so slightly that no effect was even observable.

These studies excluded five of the voting systems we have been discussing. The Dodgson, Young, and Kemeny systems were all excluded because they entail prohibitively complex computations. As mentioned in Section 8.4, that also destroys their transparency, which is also important. Range voting was excluded because of its extreme susceptibility to strategic voting; see Section 8.3.2. Recall that if a Condorcet paradox appears, the Copeland system's winner is the one who wins the largest number of their two-way races. But Darlington (2016) showed that when there is a Condorcet paradox, Copeland produces ties whenever there are just three or four candidates. And in a computer simulation in the same paper, Copeland produced ties on over half the trials whenever the number of candidates was between 5 and 10. All that makes Copeland unacceptable in my opinion, so it was excluded.

For approval voting, we assumed in these studies that each voter approves any candidate whose rating by that voter exceeds a weighted average of that voter's highest and lowest ratings, with the highest rating getting 3 times the weight of the lowest rating. By this rule the average voter approved about 3 of the 10 candidates. In the plurality runoff system, the same ratings were used for the runoff, and the runoff was a majority-rules vote between the top two candidates. For the STAR system, the computer took each voter's set of ratings, adjusted them linearly so the highest score was 5 and the lowest was 0, rounded each of these adjusted scores to the nearest integer, picked the two candidates with the highest mean rounded score, and ran a majority-rules race between those two.

The first two studies included only trials with Condorcet winners, and the last two only trials with Condorcet paradoxes. Table 1 reports the results of the first two studies, and Table 2 the last two. The first two studies compare nine non-CC systems to CC systems, which always pick the Condorcet winner in Table 1. Each of these studies included 100,000 trials. Study 1 included 100 voters in each trial, Study 2 included 1000. Studies 3 and 4 compare minimax to 11 competing systems, some with CC and some without. Parallel to the difference between Studies 1 and 2, Study 3 had 100 voters per trial and Study 4 had 1000. Thus, each of the two tables allows the reader to see the differences in results between 100 and 1000 voters per trial. Study 4 had fewer trials than Studies 1-3 because on average it took a lot more computer time to find a trial meeting the criteria of Study 4.

| Table 1. Comparing 9 non-CC systems to CC systems, in 100k trials with Condorcet winners<br>Systems are ranked by the values in the third numeric column, best systems on top. | | | | | | | | |
|---|---|---|---|---|---|---|---|---|
| | Study 1: 100 voters per trial | | | | Study 2: 1000 voters per trial | | | |
| Non-CC system | Disagree-ments with CC | Trials CC won | CC percent win | CC victory margin | Disagree-ments with CC | Trials CC won | CC percent win | CC victory margin |
| Borda | 14,722 | 10,284 | 69.9 | 5,846 | 14,455 | 12,582 | 87.0 | 10,709 |
| STAR | 19,400 | 14,502 | 74.8 | 9,604 | 18,602 | 16,771 | 90.2 | 14,940 |
| Coombs | 8,203 | 6,564 | 80.0 | 4,925 | 8,510 | 7,821 | 91.9 | 7,132 |
| Bucklin | 25,889 | 21,302 | 83.2 | 16,715 | 24,863 | 23,143 | 93.1 | 21,423 |
| Majority judgment | 26,684 | 22,405 | 84.0 | 18,126 | 10,795 | 8,974 | 83.1 | 7,153 |
| Approval | 35,191 | 31,097 | 88.4 | 27,003 | 34,000 | 32,365 | 95.2 | 30,730 |
| Hare | 45,146 | 43,152 | 95.6 | 41,158 | 48,495 | 47,524 | 98.0 | 46,553 |
| Plurality runoff | 48,471 | 46,533 | 96.0 | 44,595 | 51,501 | 50,605 | 98.3 | 49,709 |
| Plurality | 66,437 | 64,663 | 97.3 | 62,889 | 69,644 | 68,848 | 98.9 | 68,052 |

| Table 2. Comparing 11 systems to minimax (MM), in trials with Condorcet paradoxes<br>Systems are ranked by the third numeric column, yielding the same order as in Table 1. | | | | | | | | |
|---|---|---|---|---|---|---|---|---|
| | Study 3: 100k trials, 100 voters each | | | | Study 4: 10k trials, 1000 voters each | | | |
| System | Disagree-ments with MM | Trials MM won | MM percent win | MM victory margin | Disagree-ments with MM | Trials MM won | MM percent win | MM victory margin |
| Schulze | 3,965 | 2,021 | 51.0 | 77 | 46 | 25 | 54.3 | 4 |
| Black (uses Borda) | 51,899 | 27,901 | 53.8 | 3,903 | 6,068 | 3,781 | 62.3 | 1,494 |
| STAR | 57,125 | 32,328 | 56.6 | 7,531 | 6,351 | 4,108 | 64.7 | 1,865 |
| Coombs | 53,827 | 32,428 | 60.2 | 11,029 | 5,936 | 3,603 | 60.7 | 1,270 |
| Nanson | 34,445 | 21,964 | 63.8 | 9,483 | 1,449 | 932 | 64.3 | 415 |
| Bucklin | 59,753 | 38,339 | 64.2 | 16,925 | 6,598 | 4,416 | 66.9 | 2,234 |
| Majority judgment | 60,389 | 38,917 | 64.4 | 17,445 | 5,855 | 3,544 | 60.5 | 1,233 |
| Approval | 64,403 | 44,659 | 69.3 | 24,915 | 6,843 | 4,783 | 69.9 | 2,723 |
| Condorcet-Hare | 72,903 | 58,281 | 80.0 | 43,659 | 7,209 | 5,502 | 76.3 | 3,795 |
| Plurality runoff | 73,468 | 59,343 | 80.8 | 45,218 | 7,277 | 5,703 | 78.4 | 4,129 |
| Plurality | 78,023 | 66,952 | 85.8 | 55,881 | 7,827 | 6,527 | 83.4 | 5,227 |

The top line of each table names one voting system which is the "base system" for that table. The table compares every other system in the table to that system. In Table 2 the base system in minimax. In Table 1, all CC systems can be thought of as a single system, because all trials in Table 1 have Condorcet winners and all CC systems pick that person as winner, so they all pick the same winner. CC systems are the base system in Table 1.

The first numerical column for each study shows for each non-base system the number of trials in which it disagreed with the base system by picking a different winner. The next column shows the number of those trials in which the winner picked by the base system had a higher mean rating than the winner picked by the non-base system. The third column shows the percentage of disagreements won by the base system. The fourth column shows how many more trials in that study were won by the base system than by its competitor. As the tables show, those numbers were always positive, meaning the base system won more trials than its competitor. In each table, systems are arranged by the first "percent" column in that table, in declining order of performance. Happily, that rule puts all the non-CC systems in the same order in both tables. The figures in the second percent column in each table are not all in exactly that same order.

**5.2.3 Results and discussion**

We focus mainly on the two "percent" columns in each table. We see that in Study 1, CC systems picked better than every non-CC system in at least 69% of the disagreements, and in Study 2 in at least 87% of the disagreements. And CC outperforms all but one of its competitors by broader margins in Study 2 than in Study 1. The only exception to that statement is majority judgment, which gained a trivial 0.9% in Study 2, and is far below CC in both studies. Table 1 is for trials with Condorcet winners, which in real life constitute the great majority of all elections. In Table 2, majority judgment, Condorcet-Hare, plurality runoff, and plurality all gain small amounts against minimax with the larger samples, but they are far below minimax in both halves of Table 2.

Nearly every public election in the US uses either the plurality, plurality runoff, or Hare system, and Table 1 shows that when there was a Condorcet winner, all of these systems picked a worse candidate than the Condorcet winner in over 95% of the trials in which they picked different winners.

Table 2 includes four CC systems besides minimax: Schulze, Nanson, Black (which uses Borda for the trials in Table 2), and Condorcet-Hare (which uses Hare). No system ever outperforms minimax in either half of Table 2, and minimax always outperforms its competitor by at least 60-40 for all systems except Schulze, Black, and STAR. For Black and STAR, that outperformance still exceeds 60-40 for the larger sample size, which is the more important case. And Borda-Black and STAR both badly underperform minimax in Table 1.

The Condorcet-Hare system is of special interest because Green-Armytage, Tideman, and Cosman (2016) found it to be superior to minimax in resistance to strategic voting. But minimax outperforms Condorcet-Hare by over 3:1 in both halves of Table 2. Section 6 discusses this issue at length, and Section 6.3 describes another simulation study which compares the efficiency of those two systems under 32 different conditions. The results of that study were similar to the results in Table 2.

That leaves only Schulze as a possible viable competitor to minimax. There are three senses in which the two systems produce nearly equal results. First, they both pick every Condorcet winner. Second, under Condorcet paradoxes, Study 4 shows them picking different winners in only 46 of 10,000 trials, or one trial in every 217. Third, even when they do pick differently, they are essentially tied in the quality of the winners. But Schulze is far more complex than minimax in three ways – in programming the system, in computing winners once programmed, and in voter understanding of the system. Also, in small data sets, Schulze produces many more ties than minimax, which has a very effective tie-breaker described in Section 7. As mentioned above, trials with ties were eliminated from these analyses, thus making Schulze look better than it would otherwise. Thus, with equal performance, greater simplicity, and a better tie-breaker, minimax seems clearly preferable to Schulze.

These studies are new in version 8 of this paper. Several other simulation studies, somewhat different from these, can be found in version 7. They were omitted here to keep this section from being even longer, but some readers may find them interesting. They all found minimax to outperform other systems. Thus, minimax appears to offer a unique combination of efficiency (ability to pick the best winners) and simplicity.

# 6. Strategic voting under minimax

### 6.1 Some strategy-resistant features of minimax

The most-discussed form of insincere strategic voting is "burying," in which voters insincerely rank one candidate X at the very bottom, to prevent them from beating another candidate Y, even though their sincere rating of X is above that of some other candidates. There are several reasons why burying is more difficult, less effective, and more dangerous under minimax than under several other systems.

First, some systems allow what we might call a "multiplier effect" for burying. Consider the Borda system. Candidate X's Borda count is the sum of the number of candidates ranked below X by all voters. The winner is the one with the highest Borda count. Thus, if there are two major candidates A and B, and 8 minor candidates, then persuading one voter to bury A will do as much to defeat A as persuading 9 voters to rank them second after B. The mathematical details are different for each system, but Section 8 shows that the range voting, majority judgment, Coombs, Borda, and Dodgson systems all allow small numbers of voters to exercise disproportionate effects through strategic voting. Minimax has no similar effect. The minimax winner is the candidate with the lowest LL value, and even moving candidate X from one voter's top rank to their bottom rank will never raise X's LL value by more than 1.

Second, under several voting systems, strategic voting never "backfires" against the plotters, harming rather than helping the selfish interests of the strategic voters. But under minimax, even if strategic voting might work for the plotters if they guess correctly how other voters will vote, it might work against them if they guess wrong. Strategic voting typically works under minimax only if there is a Condorcet paradox in the sincere voting or if the insincere voting can create one. For example, suppose that in an election with 15 voters, 6 voters would sincerely rank three candidates ABC, 5 would rank them BAC, 2 would rank them CAB, and 2 would rank them CBA. Under sincere voting, A is the Condorcet winner and wins under all 12 of the voting systems we consider in Section 5.2. (Actually, A ties with B under the little-used Bucklin system, but Borda and plurality would be the most obvious tie-breakers, and either of those would make A win.) But suppose minimax is used, and the 5 BAC voters insincerely vote BCA. This will produce a Condorcet paradox in which A loses to C by 3 votes, C loses to B by 7 votes, and B loses to A by 1 vote, so B wins for having the smallest loss. Thus, B-voters can win by voting insincerely. But the scheme requires the plotters to know exactly how everyone else will vote. If they guess wrong, their scheme may make C win, and C is their least favorite candidate. And the scheme worked only because it created a Condorcet paradox – a rare circumstance.

Even in cases like this, schemes like this can pose difficulties and dangers for the plotters. Voters will naturally assume that sincere voting is the most effective way to promote their own interests, so it will usually take some real work to persuade voters to participate in such schemes. And because these schemes lack a multiplier effect, the schemers must persuade a lot of voters. Thus, in elections of any size, the schemes might well end up in the newspapers, well before Election Day. This will alienate independent voters, and persuade opposing voters to actually get out and vote. And the candidate promoting the scheme has been revealed as a dishonest schemer. If I vote for them, how do I know they will honor their campaign promises once they're in office?

And how often do plotters know so accurately how everyone else will vote? In elections in villages and small cities, there are often no reliable pre-election polls at all. And polls can be wrong even in large elections. In the US presidential election of 2016, some respectable pollsters predicted with 99% confidence, the day before the election, that Hillary Clinton would win. Strategic schemes under minimax require information vastly more accurate than that. Last-minute surprises can occur in elections; some are intentionally held to the last minute by candidates or their campaigns. Even the weather can offer surprises. One group might carefully plan their strategy, then see it disrupted by a last-minute surprise. The US presidential election of 2016 provides an example of both the inaccuracy of prior guesses of the outcome and of last-minute surprises (some of Hillary Clinton's work-related e-mails found on the laptop of a purported sex offender who was not her employee).

Also noteworthy is that there may be two or more conflicting plots. If A and B are major candidates and C is generally considered a minor candidate, then those favoring B will try to make C beat A as described above, and those favoring A will try to make C beat B. If both groups succeed in those goals, then C will become a Condorcet winner, and both the A and B voters will have seen their least favorite candidate elected. So, if you believe your major opponent will attempt this plot and may well succeed, then your optimum strategy is to execute no plot so that the winner will be your second-favorite candidate instead of your least-favorite candidate.

All this suggests that plots under minimax are complex and dangerous enough to discourage most of them. Thus, society should at least try minimax as an experiment. We have already seen that many of these plots won't remain secret even before voting ends, let alone after that. If it becomes obvious that some of these plots have actually succeeded, and voters everywhere start planning to try to imitate those successes despite all the complexities mentioned here, then we can accept that the experiment failed.

**6.2 Is Condorcet-Hare more strategy-resistant than minimax?**

In a major analysis, Green-Armytage, Tideman, and Cosman (2016) found only one voting system (Condorcet-Hare) which seemed to be more strategy-resistant than minimax and only slightly less efficient. When I compared the strategy resistance of these two systems in a very different computer simulation, I found results similar to theirs. Using the general procedures described in Section 5, my computer generated trials with 5 candidates and 1000 voters, all programmed to vote sincerely. The computer ran until it had found 100,000 of these trials which all met two conditions. First, each trial had to have a Condorcet winner, whom we'll call Sincere Winner or SW. Second, when SW was removed, there was a Condorcet winner among the remaining candidates; we'll call that winner SW's

Main Challenger or MC. Because minimax and Condorcet-Hare both name all Condorcet winners as winners, SW and MC would be chosen as winners, under sincere voting, by both these systems. Then, to simulate strategic voting by voters favoring MC over SW, the computer changed all the ballots of just those voters, to rank SW at the very bottom. After those changes, there were 30,298 trials in which SW remained the winner under both minimax and Condorcet-Hare, and 35,651 trials in which SW now lost under both systems. But there were 30,210 trials in which SW lost under minimax but won under Condorcet-Hare, and only 3823 trials with the opposite result. (These four frequencies sum to 100,000.) Thus, Condorcet-Hare did display far greater resistance to strategic voting than minimax, in this extreme simulation in which every relevant voter voted strategically.

But did the strategic voters actually gain from their plot? Under minimax, the strategic voters ended up electing MC in 32,725 trials, but electing some other candidate (neither SW or MC) in 33,136 other trials. This was despite the fact that under sincere voting in these artificial data, MC was the Condorcet winner in SW's absence in every single trial. The plotters did even worse in this respect under Condorcet-Hare; those two frequencies were then respectively 15,657 and 23,835. Those numbers again show Condorcet-Hare to be more strategy-resistant than minimax. But the plotters did fail in their specific goals most of the time under either system, despite having a situation tailor-made to encourage strategic voting.

Another analysis, using more data from these same 100,000 trials, added more detail to this conclusion. In our artificial data the computer "knew" each voter's exact rating of each candidate. When a strategic plot made SW lose under minimax, the program calculated the gain in the winner's mean rating *for the participants in the plot*. There were 65,861 such trials, and the mean gain for those voters was positive in just 32,725 or 49.7% of those trials. Thus, the plotters actually lost benefits slightly more often than they gained. Losses were so frequent because of the many trials, mentioned in the previous paragraph, in which the plot ended up electing neither SW nor MC, instead electing one of the other three candidates, who usually had lower ratings than either SW or MC. Such losses were even larger for Condorcet-Hare than for minimax; the plotters then succeeded in electing MC in only 39.6% of the trials in which they made SW lose. Thus again, Condorcet-Hare was even better than minimax in discouraging strategic voting. But the important point is that the schemers ended up with a net loss under either voting system, so there is no real incentive for such schemes under either system. Thus, strategic voting will likely remain rare, and society would benefit most from using the system which picks the best winners under sincere voting. Section 5.2 indicates that system is minimax.

### 6.3 More on the relative efficiency of minimax and Condorcet-Hare

In both studies in Table 2, Condorcet-Hare picked a different winner than minimax in over 70% of all trials with Condorcet paradoxes, and minimax picked the better winner in over 75% of those cases. I expanded on this finding by running another study comparing those two systems in more situations. This study included 32 blocks of trials, all under separate conditions. The number of competing candidates was 3 or 5 or 10 or 15. The number of voters was 15 or 35 or 75 or 155. The number of opinion dimensions in the spatial model was either 2 or 3. The study included a block of trials for each possible combination of these three factors, thus producing 4·4·2 or 32 blocks. In each block the computer ran until it had found 1000 trials in which minimax and Condorcet-Hare had chosen different winners. It then recorded the number of those 1000 trials in which minimax had picked a better winner (one with a higher mean rating) than Condorcet-Hare. Those 32 numbers ranged from 561 to 876, with a mean of 763 and a median of 770. Thus, minimax beat Condorcet-Hare in every single block, and on average its within-block victories were over 75%, just as in Table 2. (Version 4 of this arXiv paper reports a smaller study on this topic. A programming error was later found in that study, so it is now being replaced by the current larger study.)

## 7. Six variants and close relatives of minimax, and a tie-breaker

It seems advisable to allow tied ranks on a ballot, since some voters may genuinely have no preference between two candidates. Let the number of "participants" in each two-way race be the number of voters who expressed some preference between those two candidates. As in most ranked-choice voting systems, let's presume that a voter who fails to rank a candidate would rank that candidate below all others whom the voter did explicitly rank. Thus the "nonparticipants" in any two-way race are those who ranked the two candidates equally or omitted both. Since different two-way races will have different numbers of participants, Darlington (2016, pp. 15-16) defined minimax-P as a voting system in which we express each two-way margin of defeat as a proportion of the number of participants in that two-way race, and the winner is the candidate whose largest proportional margin of defeat is smallest. Darlington (2016, p. 16) found that in a simulation study, minimax-P cut the number of ties to a small fraction of what it had been with classic minimax, but classic minimax picked noticeably "better" (more centrist) candidates than minimax-P when neither system produced a tie. Neither of these results is surprising, the former for obvious reasons and the latter because races with more participants presumably produce more reliable results, and minimax-P ignores that fact.

Darlington (2016) suggested enjoying the advantages of both these systems by using classic minimax as the main system, with minimax-P as a tie-breaker. If minimax-P fails to break the tie, he suggested using each candidate's second-largest loss, then the proportional form of that loss, and continuing in that way through the results of all of a candidate's two-way races. In this system, wins are considered negative losses and are also included in the analysis. Thus, in the later tie-breaking stages, X might beat Y not by having a smaller loss but by having a larger win. This system should break nearly all ties. Simulation studies by Darlington (2016, p. 16) show that both parts of the tie-breaker do actually result in the selection of better (more centrist) candidates than random tie-breakers.

Darlington (2016, pp. 15-16, 33-35) also studied two other variants of minimax he called minimax-Z and minimax-L. Both are more complex than minimax-P, but neither outperformed classic minimax in simulations, so he did not end up recommending their use.

More recently, I studied two other ways of defining each candidate's largest loss LL. In the "winning votes" or WV variant of minimax, LL for candidate X is defined as the largest number of voters who voted against X in any of X's two-way races which X lost. In the "pairwise opposition" or PO version, it is defined as the largest number of voters who voted against X in any of X's two-way races, regardless of who won. Here I'll denote my "basic" form of minimax as MG, for "margins," since X's value of LL is defined as X's largest margin of loss. All three of these systems pick any Condorcet winner as winner, but they may pick different winners if there is a Condorcet paradox *and* there are tied or missing ranks.

I have run several studies comparing these three systems, and MG was the big winner in all of them. Here I'll report the results of just four of these studies. In each of these studies there were 10,000 trials with 100 voters in each trial. As in most of my studies, voters and candidates were all drawn randomly from a standardized bivariate normal distribution, and each voter was presumed to favor the candidate closest to himself or herself in that two-dimensional space, by Euclidean distance. The "true excellence" of each candidate was defined as 10 minus the candidate's mean distance from individual voters, so the computer "knew" which candidates were best.

The original Euclidean distances were computed to about 16 digits of accuracy, so no two distances were ever tied with each other. But MG, WV, and PO pick different winners only if there are tied or missing ranks, so I generated ties as follows. I first multiplied all the voter-candidate distances by 3, so they now ranged from roughly 0 to 12. Then I rounded each of these values to the nearest integer, so that many values were tied with each other. I then ran each two-way election in the usual way. Each of the three systems is compared in Table 3 with each of the others. A system "won" a trial if it picked a "better" candidate than its competitor. Neither system won a trial if the two systems picked the same winner on that trial.

**Table 3.** Relative efficiency of three versions of minimax in head-to-head comparisons among them.

| Number of candidates | MG vs WV | | MG vs PO | | WV vs PO | |
|---|---|---|---|---|---|---|
| | MG won | WV won | MG won | PO won | WV won | PO won |
| 4 | 3889 | 1351 | 4037 | 1273 | 503 | 172 |
| 6 | 3961 | 1307 | 4150 | 1250 | 587 | 265 |
| 8 | 4058 | 1196 | 4338 | 1091 | 775 | 336 |
| 10 | 4108 | 1249 | 4409 | 1145 | 876 | 370 |

The four rows of Table 3 are fairly similar, so the number of candidates made little difference. As one might guess, the last two columns of Table 3 show that WV and PO are much more similar to each other than either is to MG; WV and PO never picked different winners in more than 1250 of the 10,000 trials in each study. But WV always won more than twice as many trials as PO. The inferiority of PO is mirrored in the rest of the table, since MG always beat PO by larger margins than it beat WV. But the main conclusion from Table 3 is that MG is by far the best version, usually beating both WV and PO by over 3:1.

Those results match my own intuition. Suppose that in an election with 1000 voters, A lost to B 499 to 500, with one nonparticipant, while C lost to D 0 to 499, with 501 nonparticipants. WV and PO say that A's loss to B exceeds C's loss to D, since 500 voters voted against A while only 499 voted against C. That seems ridiculous to me.

Darlington (2016, p. 26) studied two other methods similar to minimax. In the method he named SSMD, the winner is the candidate with the Smallest Sum of Margins of Defeat. SSMD is the method which Tideman (2006) called the Simplified Dodgson method. In method SSSMD the winner is the candidate with the Smallest Sum of Squared Margins of Defeat. Using analyses like those in Section 5, Darlington compared both SSMD and SSSMD to minimax on their ability to find the "true Condorcet winner" whose identity had been hidden by either (a) random sampling error or (b) voter misperception of candidate positions. In each of these 2·2 or 4 comparisons, minimax's winning percentages were 96 or higher when either minimax or its competitor, but not both, picked correctly.

Thus, the best system known appears to be minimax, restricted to the Smith set, defining LL as a candidate's largest margin of defeat, and using our multi-stage tie-breaker.

# 8. More on the disadvantages of various systems

## 8.1 Why "simpler" ballots can actually make voting harder

Plurality elections suffer from vote-splitting and spoilers, and elections with runoffs certainly don't simplify the voting process. So arguments for "simplicity" tend to focus on approval voting. But approval ballots may make voting easier than ranked-choice ballots for the least-informed voters but more difficult for better-informed ones. If a voter ranks the candidates ABCD, the approval ballot forces them to also choose where to draw the line between "approved" and "unapproved" candidates. To do this rationally, the voter should consider whether the difference in merit between A and B exceeds that between B and C or between C and D. And for maximum impact a voter will also want to draw their line between the two candidates most likely to receive the most approvals from other voters, and therefore must guess who those two are. Thus a voter using an approval ballot must make three types of judgment not needed in a ranked-choice ballot. First, they must assess the relative sizes of differences in merit between various pairs of candidates. Second, they must guess which candidates are most likely to win. Third, they must decide how to combine these two judgments into a final ballot choice.

We should also distinguish between *requiring* and *allowing* voters to express more detail about their preferences. If a ranked-choice ballot allows tied and missing ranks (as most CC systems do), a voter can choose the complexity of their response. A voter could imitate plurality voting by putting one candidate at the top and ignoring all others. Or the voter could imitate approval voting by placing two or more candidates at the top and ignoring all others. Or they could tailor the complexity of their response in other ways, for instance by ranking their top three candidates and ignoring all others. Or they could vote *against* one or more candidates, without expressing preferences among the rest, by putting all the rest at the top but omitting the objects of their special disdain. People can be informed about all these possibilities in schools or even through television entertainment. All these ways of voting, and others, are allowed by the Condorcet Internet Voting Service (CIVS) mentioned in Sections 1 and 3. Darlington (2016, p. 14) shows how paper ranked-choice ballots can also allow a single voter to add one or more write-ins, while still allowing ballots without write-ins to be read by machine.

## 8.2 Problems with the major non-CC ranked-choice systems

The Borda, Hare, and Coombs systems appear to be the most important systems of this type. The Borda system allows tied ranks but the classic forms of the Hare and Coombs systems do not, because ties could prevent us from determining each voter's first or last choice. Other non-CC ranked-choice systems also differ from each other in this respect. There is no need to elaborate here on that point, since we argue here that all systems of this type are inferior to CC systems. For simplicity we assume throughout Section 8.2 that there are no tied ranks.

### *8.2.1 Examples of strategic voting under Borda and Coombs*

As illustrated by the Hare, Coombs, and Borda systems respectively, non-CC ranked-choice systems typically emphasize a candidate's number of high ranks, or their number of low ranks, or some statistic similar to the candidate's mean rank. The discussion of Burlington VT, in Sections 3 and 8.5, illustrate the problem with emphasizing the number of high ranks: this number is excessively influenced by the vote-splitting problem, in which even a Condorcet winner could get few top ranks because other candidates are so similar to them. But we now show that systems which emphasize average or low ranks are very susceptible to strategic maneuvers. To "bury" a candidate is to try to defeat them by insincerely giving them very low ranks. I'll call the opposite maneuver "strategic elevation." That name is my own, but the concept is very familiar to voters who use plurality voting. If a plurality voter regards candidate X highly but not at the very top, but thinks X has the best chance to beat candidates the voter really dislikes, they might strategically elevate X to the very top to increase the chance of defeating their least favorite candidates.

Section 6 showed that burying is rarely a real problem under minimax. But under Borda, suppose a voter would sincerely rank four candidates in the order ABCD, but they believe that B and C are by far the two candidates most likely to win. They prefer B to C, so they might choose to rank them as BADC on their ballot, thus strategically elevating B and burying C. Let $m_{XY}$ denote the difference between the Borda counts of candidates X and Y, with $m_{XY}$ being negative if X loses to Y. Then each step by which a voter elevates X or buries Y increases $m_{XY}$ by 1. Thus the more candidates there are, the more effective strategic voting might be. If $c$ is the number of candidates, a single voter voting strategically might increase $m_{XY}$ by the same amount as adding ($c$ – 2) voters who put X one step before Y.

Thus even a landslide could be tipped by just a few strategic voters if $c$ is high enough. For instance, suppose everyone agrees A and B are the two best candidates, and 55% prefer A to B while 45% do the opposite. Thus A wins by a landslide under a common definition of that term. But suppose there are 4 minor candidates so $c$ - 2 = 4, and 6% of the B-voters bury A. That's 2.7% of all the voters, so it's as if new B-voters had appeared in numbers equal to 4·2.7% or 10.8% of the original voters, thus tipping the election to B. The possibility of strategic voting also makes it far more difficult for every voter to plan how to vote most effectively, since they must try to guess each candidate's chance of winning. They will feel obligated to make those guesses, since they will assume the opposing voters are doing so.

The Coombs system is most susceptible to insincere strategic voting when there are several minor candidates who are all about equally popular because they're all little known. However, major-party strategists could arrange for those candidates to appear, in order to make their strategies work. Consider the case in which there are two major candidates A and B, plus ten minor candidates little known to many voters. Suppose for simplicity that there are 1000 voters, and both A and B are sincerely preferred to all minor candidates by every voter. Voters favoring A over B will correctly perceive B as the major threat to their candidate, and those favoring B will do the same for A. Suppose first that A is preferred to B by 900 voters, and B to A by the other 100, but A's voters all vote sincerely while B's all vote strategically by ranking A at the very bottom, so A gets 100 last-place ranks. If the 900 sincere voters all put the minor candidates in random order, then each minor candidate will get about 900/10 or 90 last-place ranks. Then A will likely be the first candidate eliminated, making B win despite A's enormous lead over B in sincere preferences.

Now suppose that things are as above, except that at least 100 voters prefer each of the major candidates, and at least 100 voters in each group vote strategically by ranking their main opponent at the very bottom. Each minor candidate will almost certainly get fewer than 100 last-place ranks since there are now just 800 or fewer sincere voters. Thus, A and B will likely be the first two candidates eliminated by Coombs, making the winner be someone who was not sincerely preferred to either A or B by even one voter.

***8.2.2 Is strategic voting a real danger?***

Some electoral theorists have argued that there is little danger of substantial insincere voting. But these examples show that elections can sometimes be tipped even by small numbers of insincere voters, and Section 8.3.2 gives even worse examples. In a world in which newspapers regularly discuss charges of gerrymandering, “fake news” designed to change votes, and legislatures dominated by one party changing the voting rules to make it harder for opposing voters to vote, it seems naive to design or choose voting systems on the assumption that strategic voting will never be a problem. Campaign workers could even encourage their voters to vote strategically, claiming that other parties could do it or have done it. Suppose a magazine article were to show that around the world there were several instances in which later analysis suggested that strategic voting might well have tipped an election, since a noncontroversial and reasonably qualified candidate had lost under Borda because they got many ranks at the very bottom. That article could lead to a worldwide increase in the frequency of strategic voting, which would lead to more such articles, etc. Widespread strategic voting could lead to an even broader feeling that society and government are corrupt and need to be overthrown, violently if necessary. All that can be avoided by using voting systems as resistant as possible to strategic voting.

It might be argued that strategic voting could be moderately common, but still wouldn’t matter because all sides would do it equally. But voters for one candidate might be more inclined than others to vote strategically if (a) they fear they’ll come in second if they vote sincerely, or (b) their candidate offers them some major benefit such as a tax break, or (c) their major opponent favors idealistic social programs opposed by the most selfish and Machiavellian voters, who are most likely to vote strategically, or (d) all three of those.

Borda would also encourage major political parties to employ a strategic maneuver which seems odd today because it would actually be counterproductive under plurality voting. Again suppose the two major candidates are A and B. B’s party could identify political aspirants much like B ideologically but far less experienced and less known. The party could encourage them to run as independents, perhaps even supporting them financially. Then nearly all sincere voters in both parties would rank each of those candidates below B, but many voters in B’s party would sincerely rank them ahead of A because of their ideology. That would change the Borda-count A-B difference in the direction favoring B. Thus there are actually two ways to bury a candidate: persuade your voters to insincerely put them at the bottom, or add minor candidates in such a way that your voters will do that simply by voting sincerely. If this were done often, it would make elections more burdensome for everyone by adding minor candidates.

***8.2.3 Two Borda-based attacks on CC systems***

Are there arguments for positional systems like Borda, which work entirely with each candidate’s set of ranks? An anonymous reviewer for a very prestigious academic journal used two examples to attack CC systems for conflicting with positional reasoning. But closer examination shows that both examples could actually support CC.

In the first example, 1001 voters rank 6 candidates in the order ABCDEF while the other 1000

voters rank them BCDEFA. Thus A is the Condorcet winner and the absolute winner, winning every two-way race by one vote and having just over half the first-place ranks. The reviewer considered it obvious that B should win, because every voter ranks B first or second, while nearly half the voters rank A last. B would win under Borda, or under a system in which the winner is the candidate who was placed in the top two ranks by the most voters. That conclusion seems almost indisputable with sincere voting under a valence model, but is far more questionable when we consider insincere strategic voting or a spatial model. The votes in the example are exactly what we would expect if C, D, E, and F were minor candidates sincerely ranked below A and B by all voters, and the B-voters all chose to bury A while the A-voters voted sincerely. Or even without strategic voting, this example could appear if votes were determined by a liberal-conservative scale like that mentioned in the discussion of Burlington VT in Sections 3 and 8.5, with voters distributed symmetrically on that scale, and with every voter ranking the candidates in the order of their closeness to themselves. Suppose candidates B-F are all close together, in that order with B closest to the mean, with A on the opposite side of the mean and just a bit closer to the mean than any of the others. For instance, suppose the mean is 5, A is at 4, and B-F are at 6.01, 6.02, 6.03, 6.04, and 6.05 respectively. This would produce the central features of the current example, with A beating every other candidate by very small margins but being ranked last by almost half the voters. But A should win since they're the most centrist candidate. In many elections the number of voters is large enough so that it's reasonable to assume a fairly smooth (though not necessarily symmetric) distribution of voter attitudes. But the number of candidates may be small enough that all sorts of odd distributions of candidates may appear, such as some candidates being quite similar to each other while others are more distant, as in this last example. The Borda system doesn't allow for that, as this example shows.

In the reviewer's second example, 240 voters rank three candidates as ABC, 261 give ACB, 300 give BAC, and 200 give CBA. Calculations show that among these 1001 voters, A beats B by one vote, A beats C by 601, and B beats C by 79. Thus A is the Condorcet winner and the absolute winner, and thus wins under the great majority of voting systems. Then 6 more voters appear, with two voters showing ACB, two showing CBA, and two showing BAC. By positional thinking, the three candidates are all tied among these six new voters, since each candidate appears twice in each of the three ranks. But adding these 6 voters to the previous 1001 changes the Condorcet winner from A to B, since B now beats A by one vote. A remains the Borda winner by a wide margin. The reviewer felt this exposes CC as ridiculous, since a set of votes showing a tie shouldn't change an election's result. That reasoning would also dismiss the Hare and Coombs systems, since under those systems the six new voters also switch the winner from A to B. My own guess is that if a neutral judge were asked to use common sense to choose a winner for this one case, without trying to specify a rule for all cases, they would say that C is an obvious loser, since C loses to both A and B by large margins. Thus we should choose between A and B by majority rule, so the six new voters should make B win by one vote although A should win without them. It's hard to see why that is ridiculous. Another reason to ignore this example is that all victories in it are by one vote out of 1001 or 1007, and almost any voting system might miss the best winner by one vote in a carefully designed example.

Since a reviewer for a very prominent journal presented these two examples as their main arguments for preferring positional systems over CC systems, and they were accepted as sound arguments by the journal's expert editor, I infer that there is a severe shortage of convincing arguments for that viewpoint.

## 8.3. Problems with rating-scale voting systems

### *8.3.1 Introduction and summary on rating-scale systems*

Section 8.3 is quite long, so I begin with a summary. Busy readers convinced by Section 8.3.2 might choose to skip Sections 8.3.3 and 8.3.4, though I believe they are also quite convincing.

The two best-known rating-scale voting systems are majority judgment (MJ) and range voting (RV). I criticize all rating-scale systems but focus primarily on these two. Insincere strategic voting is the biggest problem with all these systems. The strategic-voting example of Section 8.3.2 uses artificial data, but data which was designed to simulate real votes rather than to make MJ and RV look as bad as possible. Nevertheless, the example shows landslide elections being tipped if only 1.4% of voters vote strategically under MJ or only 1% under RV. Sections 8.3.3 and 8.3.4 show that under both these systems but especially under MJ, there are persuasive arguments which campaign workers can use to persuade their voters to vote insincerely by giving every candidate either the highest or lowest possible rating. I'll call that the *max-and-min* voting strategy. Even without strategic voting, Section 8.3.3 (on MJ) and Section 8.3.4 (on RV) both give examples in which B wins even though 98 of the 99 voters preferred A. I argue that cases much like these could arise in real life.

RV uses a numeric scale, often from 0 to 100, and the candidate with the highest mean wins. MJ uses a scale with non-numeric verbal labels such as "excellent" or "poor," and the candidate with the highest median wins. If the number of voters is even, the highest rating in the lower half is used as the median. If there is a tie between candidates for highest median, the ratings for each tied candidate are sorted from high to low. These sorted columns can be assembled into a matrix. In that matrix we find the row nearest the median row in which one candidate beats all others, and that candidate is the winner. If two such rows are equidistant from the median row, the one below the median row is used. These rules will break all ties unless two candidates have absolutely identical distributions of ratings.

Arguments for RV appear most fully at rangevoting.org. MJ was first published in 2007, but it is described and defended most fully by its creators Balinski and Laraki (2010). The book jacket can be seen online via Google Scholar, and is covered with effusive praise from Nobel laureates and other prominent electoral theorists. Aside from MJ's many flaws, that's the main reason Section 8.3 is so long.

Rating-scale systems are often defended as the only systems satisfying the subset choice criterion (SCC), also known as independence of irrelevant alternatives (IIA). Section 4.3 described SCC/IIA and showed why it can be dismissed; that criterion is not mentioned again in Section 8.

MJ and RV are the only well-known voting systems which do not reduce to majority rule (MR) in two-candidate races. Those races are free of most of the paradoxes and anomalies afflicting multi-candidate races. Therefore two-candidate races offer the simplest and clearest illustrations of the faults of MJ and RV. Thus we focus primarily on those races. After all, if a voting system doesn't behave reasonably even in two-candidate races, surely we shouldn't trust it to do so with more candidates – especially since a multi-candidate race can be thought of as a set of two-candidate races.

### *8.3.2 Using MJ or RV, how many insincere voters are needed to tip a landslide two-candidate election?*

I used a one-dimensional spatial model to study this question. These models are described in Sections 1 and 5.1. I divided a standard normal distribution *z* into 1000 equal-area sections, and I placed an artificial voter at each of the 999 borders between adjacent sections, thus simulating a normal distribution as closely as possible with 999 voters. I placed candidate A at the distribution's mean, with a score of 0. Thus A is the "perfect" candidate in this model since they're the most centrist possible candidate. I placed candidate B at z = 0.257. That puts 601 of the 999 voters to the left of B, so B is noticeably but not extremely off center. These placements made 551 of the 999 voters closer to A than to B, with the other 448 closer to B. A spatial model assumes every voter prefers the candidate closest to themselves, so the model has A's margin of victory under sincere MR voting at 551 – 448 or 103. That's over 10% of the voters, so with sincere voting the election is a landslide under a common definition of that term. I defined each voter's rating of each candidate as 10 minus the absolute distance between that voter and that candidate. A would win under either RV or MJ as well as under MR, since A's mean and median ratings both exceed B's.

But then, starting with the most extreme voters at the end of the distribution in which all voters favored B, I changed the rating of each voter, one at a time, to max-and-min ratings. Specifically, I had each of these voters now give A the lowest rating sincerely given to either candidate by any voter, and each give B the highest such rating. I applied the changes to the voters at the far end of the normal distribution because I judged that in real life these would be the voters who could most easily be persuaded to vote insincerely to increase B's chance of victory. Those at the extremes are often alienated from society, and thus could probably be easily persuaded to lie a bit to advance their own personal agendas. After making these changes for each voter (one voter at a time), I recomputed the two median ratings to study MJ, and stopped when B's median rose above A's. This took just 14 voters. That's just 3.1% of the 448 voters favoring B, and 1.4% of all voters. It's hard to imagine that B's campaign workers couldn't persuade that number of voters to vote strategically. When I did the same analysis using means rather than medians, thus assessing RV instead of MJ, it took only 9 insincere voters to tip the election. That's about 2% of the 448 voters preferring B, and under 1% of all voters. Thus RV is even more susceptible to strategic voting than MJ, though both are unacceptably susceptible.

Note that in both MJ and RV, all these voters were counted as rating B above A even under sincere voting; the only switch was to insincere ratings. The MR result would not be affected at all by the insincere votes. That is, if vote-counters were given MJ or RV ratings and then applied MR after noting merely which candidate each voter preferred, they would find the same winner regardless of whether voters had voted sincerely or insincerely, and the same winner as in a poll with ordinary MR ballots.

For those interested, here is more detail concerning this example. If a voter changes their rating of candidate X, it doesn't change X's median at all if the rating stays on the same side of the median. Of the 448 voters favoring B, the most centrist 198 rate both A and B above their own medians. If some of these 198 voters switch to max-and-min, under MJ it will help B by lowering A's median, but it will not raise B's. Under MJ the next most centrist 80 of those 448 voters cannot help B at all through insincere voting, since they already rate A below A's median and rate B above theirs. The 170 least centrist of the 448 B-voters rate both A and B below their own medians. If some of these 170 voters switch to max-and-min, under MJ it will help B by raising B's median, but it will not lower A's. Thus under MJ there are "only" 198 + 170 or 368 voters favoring B who can actually help B win through insincere voting. It turns out that the 198 most centrist of these 368 voters have slightly less power to tip the election than the more extreme 170, since the tip would require 16 of those 198, not 14. I consider these to be details which weaken the example's main point only slightly. A similar analysis for RV shows that the earlier paragraphs actually slightly *understate* the power of insincere voting; under RV any 9 of the 448 voters favoring B could tip the election by using max-and-min, and some sets of 8 voters could do so.

MJ's defenders might point out that MJ typically uses a rating scale with only 5 to 8 points, while this example used a scale with infinitely fine gradations. Thus in this example there is more "room" than in a typical MJ for an insincere voter to assign a rating at the very top or bottom of all ratings. But that doesn't matter under MJ, since any rating above a median will raise that median as much as a rating at the very top of the scale, and a similar effect applies at the bottom of the scale. Thus the example's criticism of MJ applies to any form of MJ as long as the extreme points on the MJ scale were above and below both candidate medians – a condition which is presumably nearly always met. When this example was modified to use a six-point rating scale, it still required only 14 insincere voters to tip the election.

This example focused on MJ and RV, but nearly any two-candidate voting system not equivalent to majority rule (MR) must be susceptible to strategic voting. MR counts equally all voters favoring a particular candidate, so nearly any system not equivalent to MR must count some of them unequally. But when voters understand that some votes count more than others, they will have every incentive to insincerely give the response which they know will maximize their vote's impact.

It also seems almost inevitable that voters will ultimately respond to these incentives. Suppose a two-candidate election is run under some system other than MR, but for general interest or because some people demanded it, authorities also report the number of voters who rated each candidate above the other. Suppose voters then see that B won even though more voters favored A. That means that those favoring B had in some sense voted more effectively than the others. Those favoring A will immediately doubt the sincerity of the B-voters, so in the next election the A-voters will almost certainly vote insincerely themselves. Soon everyone will be using the insincere max-and-min strategy. Thus all two-candidate elections will reduce to MR. And the need for insincere voting – a need created by the failure to use MR – will promote a general attitude of cynicism about politics and distrust of the government. Nationwide or worldwide publicity could make these adverse effects widespread even if they were triggered by just a few local elections.

It seems clear that any system comparing measures of central tendency, as MJ and RV do, will be at least as vulnerable to insincere voting as MJ is. Strategic voters use extreme scores, and measures of central tendency differ in the amount they are influenced by those scores. E.g., a group's midrange is the mean of the highest and lowest scores in the group; it's obviously heavily influenced by extreme scores. Of all measures of central tendency, the median is the one least influenced by extreme scores. As already mentioned, if we change any score other than the original median score, while keeping the changed score on the same side of the median, it will not change the median at all. Thus it seems likely that MJ will be less vulnerable to insincere voting than systems which use almost any other measure of central tendency. But we have already seen that MJ itself is unacceptably vulnerable to insincere voting.

### *8.3.3 Why MJ almost forces voters to vote insincerely*

I just argued that MJ may require more people voting insincerely to tip an election than any other rating-scale system. But this section shows why under MJ, it may be especially easy to persuade voters to do that. For simplicity this section continues to focus on two-candidate elections. Any candidate will prefer that all their voters use the max-and-min strategy, so campaign workers will try to persuade voters to do so. This section describes three arguments which seem particularly persuasive for MJ elections.

The "post-election impact" argument is aimed at voters who favor a particular candidate because they favor that candidate's policy positions rather than their good looks or magnetic personalities. Whether the office won is legislative or executive, a "big win" will increase the respect an office-holder receives, and it will thus help them implement their policies better than a smaller win. Therefore voters will want not merely to elect their favored candidate, but to make that candidate's win as impressive as possible. Under MJ, news media will presumably publish the number of voters rating each candidate at each level of the rating scale, and politicians will study those figures carefully. Thus the bigger the difference between the winner's figures and the loser's, the easier the winner will find it to implement their policies. In fact, if A beat B without voter X, and X rated A just slightly above B, adding X's ratings to the data set might actually lower the difference between the two sets of ratings and thus make it harder for A to implement their policies once in office, much as if X had preferred B. X doesn't want that, and the surest and easiest way for X to prevent it is to use max-and-min.

The "no impact" and "reverse impact" arguments for using max-and-min define a "single-sided" voter as one who rates both candidates above the medians of both, or rates both candidates below both medians. In a spatial model, single-sided voters will be especially numerous if the two candidates are close together in space, since voters close to both will then rate both above both medians, and voters far from both will rate both below. In a simulation study in which the spatial positions of both voters and candidates were picked randomly from a bivariate normal distribution, on the average about 40% of all voters were single-sided, and that frequency approached 99% in trials in which the two candidates were especially close to each other. That study is described more fully at the end of this subsection.

The "no impact" argument points out that if a single-sided voter were to switch their ratings, giving candidate B the rating they had previously given A and vice versa, it could not possibly change the winner because neither of the two medians would change at all. So their preference between candidates has no effect at all on who wins. The "reverse impact" argument notes that a single-sided voter's decision *to* vote could actually make their preferred candidate lose. For instance, suppose 4 voters voted before X did, and their ratings were 1 2 5 6 for candidate A and 1 3 4 6 for B. When the number of voters is even, MJ uses as a candidate's median the lower of their two central ratings. Thus in this example, B was winning because the calculated medians were 2 for A and 3 for B. Suppose X then votes, rating A at 5 and B at 6. This moves the medians up to 5 and 4 respectively, making A win even though X prefers B and B was winning before X voted. This illustrates MJ's nonmonotonicity, which was mentioned in Section 3.

In one sense the "reverse impact" argument is weaker than the other two, because it requires an exceptionally large gap between adjacent ratings for one candidate. But the real question is whether these arguments could be used to persuade voters to use max-and-min, and I think all three could be so used. Thus a campaign worker might say, "If you don't use max-and-min, your vote might not really be counted, or might lower the published win size of your favored candidate, or might even make them lose. And max-and-min actually requires less hard thinking than the alternative."

The anomalies of single-sidedness can be pushed to an extreme. For instance, suppose ratings are on a scale from 1 to 6, with 6 high. Suppose 49 voters rate A at 2 and B at 1, 49 others rate A at 6 and B at 5, and one voter rates A at 3 and B at 4. Then A's median is 3 and B's is 4, so B wins under MJ even though 98 of the 99 voters prefer A to B. Those 98 voters are all single-sided, so their preferences don't count. Balinski and Laraki (2010, pp. 328-329) dismiss examples like this as contrived and unlikely to occur in real life. But in the simulations mentioned above and described more fully in the next few paragraphs, the number of voters whose preferences didn't count did approach 99% in some trials designed to simulate real elections. And this paragraph's 99-voter example merely extends to extremes the problems for single-sided voters which affect virtually every use of MJ. Note that the anomaly in this paragraph does not require insincere voting.

The aforementioned computer simulation was as follows. I used a two-dimensional spatial model with a bivariate normal distribution of voters. I drew a random sample of 10,000 voters and used that same sample for all of the 100,000 trials about to be described, since I assumed that single set of 10,000 voters provided an adequate approximation to a bivariate normal population. Then I randomly drew 100,000 pairs of candidates from the same bivariate distribution; each such pair constituted a new trial. In each trial I computed the Euclidian distance between each candidate and each voter, and defined the voter's rating of that candidate as 10 minus that distance. I then computed each candidate's median rating, and used those medians to label each voter as single-sided or not. The mean number of single-sided voters per trial (out of 10,000 voters in the trial) was 4004.7, and the median was 3929. Thus we can say that on average, about 40% of voters were single-sided. But as mentioned four paragraphs ago, those numbers were higher when two candidates were similar and lower when they were dissimilar. Thus the number of single-sided voters in a trial ranged from 0 to 9889, so there were trials in which about 99% of voters were single-sided.

I also defined a voter's "preference intensity" as the absolute value of that voter's difference between the ratings of the two candidates. On average the intensity of single-sided voters was about 66% of the intensity of other voters. Across all trials, about 32% of the total voter intensity was found in single-sided voters. But there were trials in which over 99% of the total voter intensity was in single-sided voters, since that percentage correlated -0.81 with the Euclidian distance between the two candidates and those distances ranged from under 0.01 to 6.87.

In the 100,000 trials, there were just 1128 trials in which MJ and MR (majority rule) picked different winners. Each system picked the more centrist of the two candidates on over 90% of the trials. But when the two systems picked different winners, MR picked the more centrist candidate on 840 trials and MJ did so only on 288. Thus MR "won" on 74% of these trials. That difference is enough to persuade a neutral outsider that MR is superior, even assuming everyone votes sincerely. It also turned out that the more single-sided voters there were in a trial, the more likely MJ was to pick the "wrong" (less centrist) candidate in that trial. And the results in this subsection seem more than enough to persuade many voters to vote insincerely in order to avoid being classified as single-sided by MJ.

Darlington (2017) describes several other simulation studies showing how poor MJ is at picking the best candidates, even under sincere voting.

#### *8.3.4 Like MJ, RV can totally ignore majority rule*

Suppose we have a rating scale from 0 to 100. Suppose there are 99 voters, and 98 of them rate A one point above B, but the $99^{th}$ voter rates A at 0 and B at 100. Then B's rating total across voters exceeds A's by two points. Thus B wins by RV although 98 of the 99 voters preferred A to B. B might seem like a reasonable winner in this example until we consider the possibility of max-and-min strategic voting. A single strategic voter who favors B would vote exactly as in this illustration, thus tipping an election in which every other voter prefers A.

Campaign workers would not find it as easy under RV as under MJ to persuade their voters that they *should* vote strategically, since there is no obvious sense in which RV ignores anyone's sincere vote. But other arguments for strategic voting still apply, such as the desire to make your candidate's win as big as possible, and the fear that opposing voters may vote strategically and may have done so in the past. And as we saw in Section 8.3.2, it usually takes fewer insincere voters to tip an election under RV than under MJ. One reason for this is that if a voter favoring B sincerely rates B above B's median and sincerely rates A below A's median, strategic voting by that voter will not help B at all under MJ. But under RV every strategic vote helps the favored candidate, except for the rare voter who sincerely rates their favored candidate at the very top and their opponent at the very bottom. Thus overall, it's hard to argue that RV is more resistant than MJ to strategic voting.

## 8.4 Problems with the Dodgson and Young systems

Aside from minimax, Dodgson and Young are the only well-known voting systems which satisfy both CC and SDVC. This section describes their limitations. I know of no countervailing advantages of these systems over minimax.

Oxford mathematics lecturer Charles Dodgson is better known as Lewis Carroll, the author of *Alice's Adventures in Wonderland*. In a pamphlet printed in 1876 and reprinted in Black (1958), Dodgson considered a voter who interchanges two candidates whom the voter had previously ranked adjacently, as when the ABCD ranking is changed to ACBD or ABDC. Reversing ABCD to DCBA would require six such interchanges. Dodgson suggested naming as winner the candidate who would need the smallest number of such interchanges, summed across voters, to become a Condorcet winner. In a different approach, Young (1977) considered voters who help defeat a candidate by ranking them poorly. He suggested counting for each candidate X the smallest number of such voters who would have to be deleted to turn X into a Condorcet winner, and naming as winner the candidate for whom this number was smallest.

Perez (2001) showed that both the Dodgson and Young systems lack monotonicity. The important writers Felsenthal and Nurmi (2017) consider monotonicity essential, and I agree. Nonmonotonicity also makes a system susceptible to strategic voting, and Green-Armytage (2014) found minimax to be one of the systems most resistant to strategic voting.

The conceptual simplicity of the Dodgson and Young systems also masks substantial computational difficulty. Hemaspaandra, Hemaspaandra, and Rothe (1997) show that the Dodgson system is so difficult it can be a challenge even for modern computers. Rothe, Spakowski, and Vogel (2003) reach a similar conclusion for the Young system. Even if those difficulties can be overcome, they raise an important issue of transparency. A voting system is transparent if the average citizen can see that it would be difficult or impossible for corrupt insiders to alter the election's results. But under Dodgson or Young, there is no way an average citizen could understand or check the necessary computer programs, meaning they would have to accept their conclusions on faith. That can be a major political problem when citizens don't trust governmental institutions, as is widely true. In contrast, minimax would use a computer program which counts the number of voters who rank one candidate X above some other candidate Y, and does that for every pair of candidates. That program would be so simple that the program writers would have no plausible claim that the program must remain confidential to protect trade secrets, and even a small town would contain people who could check the honesty of such a program. Once that step is done, the remaining calculations in minimax are so simple they could easily be done by hand by almost anyone: we compute all the margins of victory and defeat, we find each candidate's largest margin of defeat, and name as winner the candidate for whom this margin is smallest.

Some European governments have attempted to maximize transparency by publishing replicas of all ballots. But according to Naish (2013), the Italian mafia appears to have devised a way to use this information to force many voters to vote as the mafia wishes, at least in elections with many candidates. They can direct the voter to put the mafia candidate first, but direct them to use the next few ranks to endorse a set of candidates so opposed to each other that no sincere voter would ever choose that set. Those ranks constitute the voter's "signature" which allows the mafia to verify that the voter put their candidate first. We have just seen that minimax can achieve a very high level of transparency without sacrificing voter privacy. But that's not true for the Dodgson and Young systems, since an outsider couldn't really check every step without seeing each voter's actual ranking of candidates, thus running into the Naish problem. The tie-breaker recommended in Section 7 uses the same vote totals used by simple minimax, and thus constitutes no greater threat to voter privacy.

The Dodgson system can also make it easy to "bury" a candidate – defeat them by having voters insincerely put them at or near the bottom of their rankings. For instance, suppose we have 25 voters and 6 candidates A-F, with the following circumstances:

1. All voters perceive A, B, and C as the best candidates and the ones most likely to win. Under sincere voting everyone would put D, E, and F at the bottom in that order.

2. It's generally agreed that the most likely winners are A, B, and C in that order. Thus voters favoring C will try to bury A and B, and voters favoring B will try to bury A.

3. A Condorcet paradox appears among these three candidates, with A beating B in their two-way race, B beating C, and C beating A. A Condorcet paradox is an essential feature of any example comparing CC systems, since without the paradox, all those systems would pick the same winner.

Consistent with all these specifications, we have:

| Pattern | Frequency |
|---|---|
| ABCDEF | 10 |
| BCDEFA | 8 |
| CDEFAB | 7 |

The numbers of Dodgson interchanges needed to make A, B, or C a Condorcet winner are 12, 5, and 6 respectively. Thus A loses big, with at least twice as many interchanges needed as for either B or C. But under minimax the largest two-way losses of the six candidates, in alphabetical order, are 5, 9, 11, 25, 25, and 25, so A beats even B and C by wide margins relative to the number of voters. This difference in outcomes is caused entirely by the presence of candidates D, E, and F and the burying strategies used against A and B. But D, E, and F are all losers by any reasonable voting system, since they are all ranked below C by every single voter. If we remove them from the race, the largest two-way losses of A, B, and C remain the same, so A still wins under minimax. But their Dodgson values become 3, 5, and 7 respectively, so A now wins under Dodgson as well. With D, E, and F gone, A also wins under the Borda, Hare, Coombs, plurality, and many other systems, as well as under minimax and Dodgson. It thus seems clear that A is the best candidate, though Dodgson put A far behind B and C when all 6 candidates were included. And this is with just three minor candidates (candidates who would lose big by virtually any rule). The more minor candidates there are, the more "dirt" each insincere voter has to pile onto any candidate they wish to bury, and the fewer such voters it takes to bury a candidate.

As mentioned above, Young (1977) proposed naming as winner the candidate with the smallest number of voters who would have to be removed to turn them into a Condorcet winner. He commented (p. 350) that the Young and minimax systems are probably very similar, and I agree. But if anything, minimax is superior to Young when we think more carefully about how votes would need to change to turn a candidate into a Condorcet winner. Since a Condorcet winner is defined in terms of two-way races, consider the three ways a voter could act more favorably toward B in a two-way race between A and B which A had won. (1) The voter could abstain when they had previously favored A. (2) They could favor B when they had previously abstained. (3) They could favor B when they had previously favored A. There are three parallel ways they could act more favorably toward A than they had before, making 6 ways a voter could change. We would ideally like to consider the effects of all of those possible changes. If we did so, we would presumably want to add them up, presumably giving (3) twice the weight given to (1) and (2). The three parallel ways of changing toward A would be given negative weights, with the third again being given twice the weight of the others. The Young system considers just the first of those six types of possible change. One's first thought is that it would be wonderful but completely impractical to consider all six of these types of change. But surprisingly, it's actually much easier to consider all six of these together, using the additive rule just described, than to implement the Young system. That's because the additive rule simply yields the change in B's margin of defeat against A. Thus we can effectively implement that rule by using minimax, in which the winner is the candidate whose margins of defeat have the smallest maximum.

We can also say that the minimax winner is the candidate X who would need the fewest new voters, all putting X first, to transform X into a Condorcet winner, since that number is always just one greater than X's largest two-way margin of defeat. But that statement about new voters doesn't mean that minimax is ignoring the other types of voting change just mentioned; it just means that those other types of change can be transformed into numbers of new voters needed by the additive rule of the previous paragraph.

### 8.5 Problems with the Hare system (instant runoff voting or single transferable vote)

The Hare system is by far the best-known ranked-choice voting system, and has been used for public elections in at least eight English-speaking countries, though in few others. A serious problem with that system is the frequency with which a Condorcet winner can lose despite winning all their two-way races by majority rule. This occurred one of the first times the Hare system was used for a major public election in the US. That was in 2009, when it was used to elect a mayor for Burlington, VT. The Democratic candidate won all five of his two-way races, but was eliminated by Hare before either the Republican or the Progressive candidate, who won. In the two-way race between them, the Democrat actually beat the Progressive by a 7.8% margin; a 10% margin is sometimes called a landslide. The voters of Burlington found the Hare choice bizarre, and voted shortly thereafter to discard that system. This anomaly can never occur under minimax.

To see how this can happen under Hare, suppose voters and candidates differ primarily on a liberal-conservative or left-right dimension, A is slightly left of center, B is very centrist, and C is slightly to B's right. Then in a standard spatial model, A would receive top ranks from all voters to the left of the midpoint between A and B, and C would receive top ranks from all those to the right of the midpoint between B and C, while B receives top ranks only from the few remaining voters. Thus, B might well receive the fewest top ranks even though B would beat either A or C in a two-way race.

I ran a computer simulation to estimate how often this might happen. The number of candidates in a trial was 3, 5, 10, or 15, and the number of voters was 15, 35, 75, 155, or 501. A spatial model was assumed to have 1, 2, or 3 dimensions. Each possible combination of these three factors was studied, producing 4·5·3 or 60 different conditions, with 10,000 trials in each condition. Less than 2.5% of these 600,000 trials had Condorcet paradoxes, so nearly all had Condorcet winners. Call it an HC anomaly if the Hare and Condorcet winners are different people. It turned out that the number of voters per trial had little effect on the frequency of these anomalies, so I'll report the results for just the largest number of voters per trial (501). Table 4 shows these frequencies. We see that the frequency of HC anomalies depends enormously on both the number of candidates and the number of opinion dimensions assumed. In real life both these factors will vary greatly across elections.

I'll discuss the results in Table 4 after mentioning another feature of this study. As mentioned above, Table 4 gives results for just 12 of the 60 conditions studied, because each of the 12 conditions in Table 4 was also replicated with 15, 35, 75, or 155 voters per trial. In each of these 60 conditions I also computed the proportion of times that the Hare winner was inferior to (less centrist than) the Condorcet winner when those two were different people. Those 60 proportions ranged from 0.808 to 0.979, with a mean of 0.921 and median of 0.929. That supports a conclusion which I believe most people would accept intuitively – when the Hare winner and Condorcet winner are different people, the Condorcet winner should clearly win. Tables 1 and 2 in Section 5.2 suggests the same conclusion.

I'll use that fact in interpreting the results of Table 4. We see that in one of the conditions there, the HC anomaly appears in over three-fourths of all trials. And even in the condition in which these anomalies are rarest, they would be expected to occur about once in every 47 elections (213 times in every 10,000 elections). If we assume (as I do) that even a single occurrence of this anomaly would be a substantial embarrassment for proponents of Hare, these results offer a strong reason to avoid Hare.

**Table 4.** Number of HC anomalies (trials in which the Hare winner and Condorcet winner were different people), in 10,000 artificial-data trials in each of 12 conditions, with 501 voters in each trial. Here *c* is the number of candidates in each trial, and *d* is the number of opinion dimensions assumed.

| **c\d** | **1** | **2** | **3** |
|---|---|---|---|
| **3** | 1506 | 409 | 213 |
| **5** | 4090 | 1621 | 691 |
| **10** | 6863 | 4797 | 2379 |
| **15** | 7769 | 6780 | 4302 |